\documentclass[pra,twocolumn,showpacs]{revtex4}
\usepackage{graphicx}
\usepackage{amsmath}
\usepackage{amssymb}

\def\lsim{\
  \lower-1.2pt\vbox{\hbox{\rlap{$<$}\lower5pt\vbox{\hbox{$\sim$}}}}\ }
\def\gsim{\
  \lower-1.2pt\vbox{\hbox{\rlap{$>$}\lower5pt\vbox{\hbox{$\sim$}}}}\ }

\begin{document}

\title{Expansions of the interatomic potential for different
       boundary conditions \\ and the transition to the thermodynamic limit}

\author{Maksim D. Tomchenko}
\affiliation{Bogolyubov Institute for Theoretical Physics,
        14-b, Metrolohichna Str., Kyiv 03143, Ukraine}

\date{\today}
\begin{abstract}
We analyze the possible expansions of the interatomic potential
$U(|\textbf{r}_{1}-\textbf{r}_{2}|)$ in a Fourier series for a
cyclic system and a system with boundaries. We also study the
transition from exact expansions for a finite system to the
expansion that is commonly used in the thermodynamic limit. The
analysis shows that such a transition distorts the potential of a
bounded system by making it cyclic.

       \end{abstract}


\maketitle

\section{Introduction}
\noindent The theoretical studies of many-particle problems
frequently use
 the expansion of the interatomic potential
$U(|\textbf{r}_{1}-\textbf{r}_{2}|)$ in a Fourier series. In this
case, the real-world finite system is replaced by an infinite one:
the periodic boundary conditions are used, and it is assumed in the
formulae that the system sizes and the number of particles tend to
infinity at the constant ratio $N/V$. This procedure is called the
transition to the thermodynamic (T-) limit, and it gives the
below-presented expansion (\ref{1})-(\ref{3}). The expansion
(\ref{1})-(\ref{3}) is used throughout the literature on condensed
matter and in statistical physics as the standard trick. However, we
have not been able to find in the literature the mathematical
justification that the potential of a finite bounded system can be
expanded as the potential of an infinite system. In one of few books
\cite{bogol1946}, where the T-transition is explained, we can read
in section 7: ``In our entire consideration, ... the limiting
transition $N\rightarrow \infty $ is performed purely formally. We
do not consider the complicated mathematical problem concerning the
formulation of those conditions that should be imposed on the
initial data (e.g., on the form of a potential function $\Phi(r)$)
in order to ensure the performing of a strict mathematical
substantiation of the legitimacy of such limiting transition.'' In
\cite{ruelle}, the T-transition $N, V\rightarrow \infty $,
$N/V=const$ is substantiated by means of the consideration of a
statistical sum, but no proof of the coincidence of the limits for a
cyclic system and a system with boundaries was given. Statistical
arguments in favour of the possibility of replacing real boundary
conditions by periodic ones are given in book \cite{peierls}. On the
whole, however, the validity of the transition to the T-limit for a
bounded system is in fact only a plausible conjecture. In this case,
it is assumed that either expansion (\ref{1})-(\ref{3}) is exact or
that the inaccuracy is insignificant and cannot lead to errors in
the description of a real physical system with boundaries.

The question of the legitimacy of using the T-limit is a complicated
mathematical problem. In what follows, we do not attempt to consider
it in its entirety. We only note that using of the T-limit to
describe a finite bounded system implies a change in the topology of
the system. For the classical systems, the use of
(\ref{1})-(\ref{3}) does not lead to errors and therefore seems
justified. In quantum physics, when studying the solutions of the
Schr\"{o}dinger equation, the transition to the T-limit means first
of all a change in the Fourier expansion of an interatomic potential
$U(|\textbf{r}_{j}-\textbf{r}_{l}|)$. However, when describing a
finite system with boundaries,  it is necessary to use a
corresponding expansion of the potential
\cite{mtgp2012,mtmethodbog}. Therefore, in what follows we will
analyze the possible expansions of the potential
$U(|\textbf{r}_{1}-\textbf{r}_{2}|)$ in a Fourier series for
different boundary conditions and try to determine the accuracy of
the expansion (\ref{1})-(\ref{3}).

\section{Analysis of the expansions of a potential}
The standard expansion (we will call it T-expansion) of an
interatomic potential in the T-limit reads
\begin{equation}
 U(|\textbf{r}_{1}-\textbf{r}_{2}|) = \frac{1}{V}
 \sum\limits_{\textbf{k}}^{(2\pi)}\nu(\textbf{k})e^{i\textbf{k}(\textbf{r}_{1}-\textbf{r}_{2})},
     \label{1} \end{equation}
 \begin{equation}
 \nu(\textbf{k})  =\int\limits_{-L_{x}}^{L_{x}}dx \int\limits_{-L_{y}}^{L_{y}}dy
 \int\limits_{-L_{z}}^{L_{z}}dz U(r)e^{-i\textbf{k}\textbf{r}}\equiv \nu(k).
    \label{2} \end{equation}
The symbol $(2\pi)$ above the sum (\ref{1}) indicates that
$\textbf{k}$ runs the values
 \begin{equation}
  \textbf{k}=2\pi \left (\frac{j_{x}}{L_{x}}, \frac{j_{y}}{L_{y}},
 \frac{j_{z}}{L_{z}} \right ),
    \label{3} \end{equation}
where $j_{x}, j_{x}, j_{x}$ are integers,  $L_{x}, L_{y}, L_{z}$ are
the system sizes, and $V=L_{x}L_{y}L_{z}$. In (\ref{1})-(\ref{3})
the T-limit is assumed: $L_{x}, L_{y}, L_{z} \rightarrow \infty$.

We can try to obtain T-expansion (\ref{1})-(\ref{3}) from the integral Fourier transformations \cite{kurant1}
(for simplicity, we consider the one-dimensional (1D) case, $L_{x}=L$):
\begin{equation}
f(t) = \frac{1}{2\pi}\int\limits_{-\infty}^{\infty}\Omega(\omega)e^{i\omega t}d \omega,
     \label{4} \end{equation}
 \begin{equation}
\Omega(\omega) = \int\limits_{-\infty}^{\infty}dt
  f(t)e^{-i\omega t}.
    \label{5} \end{equation}
Let us rename: $\omega = k/k_{0}$, $\Omega (k/k_{0})=\nu(k)$,
$t=x/L$, $f(x/L)=LU(x)$, where $k_{0}=1/L$, $L \rightarrow \infty$.
We also replace in (\ref{4}) $x \rightarrow x_{1}-x_{2}$  and
$\frac{L}{2\pi}\int_{-\infty}^{\infty}dk \rightarrow
\sum_{k_{j}=2\pi j/L}$. Then relations (\ref{4}) and (\ref{5}) yield
(\ref{1})-(\ref{3}). In this case, it is also necessary to
substantiate the transition from the infinite system to a finite
one.  Below, we will give a more rigirous derivation of the
T-expansion based on the consideration of a finite system.

Consider a system of $N$ particles in a finite volume $x \in [0,
L_{x}]$, $y \in [0, L_{y}]$, $z \in [0, L_{z}]$, $L_{x}\cdot
L_{y}\cdot L_{z}= V$. The function $F(\textbf{r}_{1},\textbf{r}_{2})
=U(|\textbf{r}_{1}-\textbf{r}_{2}|)$ can be expanded in a Fourier
series in three ways: i) by considering the vector $
\textbf{r}_{1}-\textbf{r}_{2}$ to be the argument (the modulus is a
part of the function); it is the zero-expansion; ii) using the
vector $(|x_{1}-x_{2}|,|y_{1}-y_{2}|,|z_{1}-z_{2}|)$ as the
argument:  the modulus-expansion; and iii) by taking
$\textbf{r}_{1}$ and $\textbf{r}_{2}$ as the arguments (in this
case, the function includes a potential, a modulus and the
difference of $\textbf{r}_{1}$ and $\textbf{r}_{2}$). One can also
invent an infinite number of unphysical expansions. They are proper,
but it is difficult to apply them to physical models. For example,
we may set $x_{1}-x_{2}= x_{1}+x_{2}-2x_{2}$ in the 1D case and
consider $x_{1}+x_{2}$ and $2x_{2}$ as independent arguments; other
combinations can be written as well. We omit these expansions.

For the three expansions considered above, we have $x_{1}-x_{2}\in
[-L_{x},L_{x}]$, $y_{1}- y_{2}\in [-L_{y},L_{y}]$, $z_{1}-z_{2}\in
[-L_{z},L_{z}]$ and $|x_{1}-x_{2}|\in [0,L_{x}]$,  $|y_{1}-
y_{2}|\in [0,L_{y}]$, $|z_{1}-z_{2}|\in [0,L_{z}]$. By the
Fourier-analysis rules \cite{kurant1}, we obtain i) Zero-expansion:
\begin{equation}
 U(|\textbf{r}_{1}-\textbf{r}_{2}|) = \frac{1}{2^{f}V}
 \sum\limits_{\textbf{k}}^{(\pi)}\nu_{z}(\textbf{k})e^{i\textbf{k}(\textbf{r}_{1}-\textbf{r}_{2})},
     \label{16} \end{equation}
  \begin{equation}
 \nu_{z}(\textbf{k}) = \int\limits_{-L_{x}}^{L_{x}}dx \int\limits_{-L_{y}}^{L_{y}}dy
 \int\limits_{-L_{z}}^{L_{z}}dz U(r)e^{-i\textbf{k}\textbf{r}} \equiv \nu_{z}(k),
    \label{17} \end{equation}
where $f$ is the dimensionality of the system, and $(\pi)$ above the
sum indicates that $\textbf{k}$ runs the values
 \begin{equation}
  \textbf{k}=\pi \left (\frac{j_{x}}{L_{x}}, \frac{j_{y}}{L_{y}},
 \frac{j_{z}}{L_{z}} \right ).
    \label{18} \end{equation}
ii) Modulus-expansion:
  \begin{equation}
 U(|\textbf{r}_{1}-\textbf{r}_{2}|) =
 \sum\limits_{\textbf{k}}^{(2\pi)}\frac{\nu_{m}}{V}(\textbf{k})e^{ik_{x}|x_{1}-x_{2}|+ik_{y}|y_{1}-y_{2}|+ik_{z}|z_{1}-z_{2}|},
     \label{19} \end{equation}
  \begin{equation}
 \nu_{m}(\textbf{k}) = \int\limits_{0}^{L_{x}}dx
 \int\limits_{0}^{L_{y}}dy \int\limits_{0}^{L_{z}}dz U(r)e^{-i\textbf{k}\textbf{r}}.
    \label{20} \end{equation}
Expansion (iii) will be considered below. Let us see the accuracy of
the above expansions using two 1D examples.

Example (A) involves the expansion of a linear potential
\begin{equation}
 U(|x_{1}-x_{2}|) =
   U_{0}|x_{1}-x_{2}|/L
\label{32}  \end{equation}
in the domain $x_{1}, x_{2}\in [0,L]$.
In this case, the zero-series (\ref{16}), (\ref{17}) takes the form
\begin{eqnarray}
 &&U(|x_{1}-x_{2}|) =
   \frac{U_{0}}{2}-\nonumber \\
 &-&\frac{4U_{0}}{\pi^{2}}\sum\limits_{j=0, 1, 2, \ldots}\frac{\cos{\left [
 \pi (2j+1)(x_{1}-x_{2})/L\right ]}}{(2j+1)^2}
\label{33}  \end{eqnarray}
and reproduces  (\ref{32}) exactly, in the required domain.
The modulus-series (\ref{19}), (\ref{20}) can be written as
\begin{equation}
 U(|x_{1}-x_{2}|) =
 \frac{U_{0}}{2}-\frac{U_{0}}{\pi}\sum\limits_{j=1,2,\ldots}\frac{\sin{[2\pi j|x_{1}-x_{2}|/L]}}{j}.
\label{34}  \end{equation} It reproduces (\ref{32}) exactly as well.
Consider the T-expansion. It turns out that in this case $\nu(k \neq
0)=0,$ and the series is reduced to
\begin{equation}
 U(|x_{1}-x_{2}|) = \nu(0)/L=   U_{0}.
\label{35}  \end{equation} In other words, the T-expansion strongly
distorts the initial potential. This occurs due to the addition of
the ``image,'' namely the potential $U(L-|x_{1}-x_{2}|)$ (see
below).

Let us consider example (B) with the ``semi-transparent sphere''
potential
\begin{equation}
 U(|x_{1}-x_{2}|) =
\left [ \begin{array}{ccc}
    U_{0}>0,  & \   0\leq |x_{1}-x_{2}|\leq a,  & \\
        0,  & \ a <|x_{1}-x_{2}|\leq L. &
\label{22} \end{array} \right. \end{equation}
In this case, the zero-series reads
\begin{eqnarray}
&& U(|x_{1}-x_{2}|) =
   \frac{aU_{0}}{L}+\nonumber \\
&+&\frac{2U_{0}}{\pi}\sum\limits_{j=1, 2, 3, \ldots}\frac{\sin{(\pi j a/L)}}{j}\cos{ [
 \pi j(x_{1}-x_{2})/L ]}
\label{p1}  \end{eqnarray}
and reproduces  function (\ref{22}) in the domain $x_{1}, x_{2}\in [0,L]$ exactly.
The modulus-series takes the form
\begin{eqnarray}
&& U(|x_{1}-x_{2}|) =
   \frac{aU_{0}}{L}+\nonumber \\
 &+&\frac{U_{0}}{\pi}\sum\limits_{j=1, 2, 3, \ldots}\frac{1}{j}\left \{ \sin{(2\pi j a/L)}\cos{ [
 2\pi j(x_{1}-x_{2})/L ]}+ \right. \nonumber \\ &+& \left.(1-\cos{(2\pi j a/L)})\sin{ [
 2\pi j|x_{1}-x_{2}|/L ]}\right \}
\label{p2}  \end{eqnarray} and also reproduces potential (\ref{22})
exactly. The T-expansion is
\begin{eqnarray}
 && U(|x_{1}-x_{2}|) =
   \frac{2aU_{0}}{L}+ \label{p3} \\
 &+&\frac{2U_{0}}{\pi}\sum\limits_{j=1, 2, 3, \ldots}\frac{\sin{(2\pi j a/L)}}{j}\cos{ [
 2\pi j(x_{1}-x_{2})/L ]}
  \nonumber \end{eqnarray}
 and yields the function
\begin{equation}
 U^{T}(x) =
\left [ \begin{array}{ccc}
    U_{0},  & \   0\leq |x|\leq a,  & \\
        0,  & \ a <|x|\leq L-a, & \\
    U_{0},  & \   L-a < |x|\leq L,  &
\label{29} \end{array} \right. \end{equation} which contains, in
addition to the initial potential, its ``image'': the same potential
at the end of the interval, i.e. at $]L-a < x\leq L]$. We note that,
at the points of discontinuity of the function, the expansion gives
the arithmetic mean of the values of the function on the left and on
the right from the discontinuity, but we omit this feature for
simplicity. We summed all series numerically. Series (\ref{33}) and
(\ref{34}) can be summed analytically at the points $x_{1}-x_{2}=0;
L/4; L/2; 3L/4; L$.

The main point is as follows. If we would expand the function
$f(x),$ which depends on the single argument and is given on the
interval $x\in [0,L],$ in a Fourier series, this would yield the
function $\tilde{f}(x)$, which coincides with $f(x)$ on the interval
$x\in [0,L]$ and is periodic with period $L$ outside the interval
$[0,L]$. In this case, $\tilde{f}(x)$ would not contain the image
inside the interval $[0,L]$: for the example B, the second ``step''
would start at the point $L$, rather than at $L-a$. In this case,
the series restores the function exactly inside the interval $[0,L]$
(except for, possibly, the end points $x=0$ and $x=L$). However, the
T-series has generated the image $U(L-|x_{1}-x_{2}|)$
\textit{inside} the interval $x\in [0,L]$. Therefore, the series
reproduces the function inaccurately. This means, obviously, the
simple point: the T-expansion is not the Fourier-expansion for a
system with boundaries. Apparently, this property has not been
noticed before. In the literature, the T-expansion is called the
Fourier-expansion, implying an infinite system, but is applied
namely to finite systems. Below, we will study in detail why the
image appears. For clearness, we have started above with simple
examples, which can be easily verified.

We will now consider the images more profoundly and give another way
of deriving the T-expansion. We expand $U(|x_{1}-x_{2}|)$ in a
Fourier series, by considering $x_{1}$ and $x_{2}$ as arguments (in
1D):
\begin{equation}
 U(|x_{1}-x_{2}|) = \frac{1}{L^{2}_{x}}
 \sum\limits_{k_{j_{1}}k_{j_{2}}}^{(2\pi)}\nu_{2}(k_{j_{1}},k_{j_{2}})e^{ik_{j_{1}}x_{1}+ik_{j_{2}}x_{2}},
     \label{6} \end{equation}
  \begin{equation}
 \nu_{2}(k_{j_{1}},k_{j_{2}}) = \int\limits_{0}^{L_{x}}dx_{1}
 \int\limits_{0}^{L_{x}}dx_{2} U(|x_{1}-x_{2}|)e^{-ik_{j_{1}}x_{1}-ik_{j_{2}}x_{2}}.
    \label{7} \end{equation}
In (\ref{7}) we make change $x_{1}=\tilde{x}_{1}+x_{2}$. Then
 \begin{equation}
 \nu_{2}(k_{j_{1}},k_{j_{2}}) = \int\limits_{0}^{L_{x}}dx_{2}
F(k_{j_{1}},x_{2})e^{-i(k_{j_{1}}+k_{j_{2}})x_{2}},
    \label{8} \end{equation}
where
 \begin{equation}
F(k_{j_{1}},x_{2}) = \int\limits_{-x_{2}}^{L_{x}-x_{2}}d\tilde{x}_{1}
U(|\tilde{x}_{1}|)e^{-ik_{j_{1}}\tilde{x}_{1}}.
    \label{9} \end{equation}
For a cyclic system,
\begin{equation}
U(x_{1},x_{2})=U(x_{1}+jL_{x},x_{2})=U(x_{1},x_{2}+lL_{x}),
\label{10} \end{equation}%
$j$ and $l$ are integers. Therefore, $U(|-x|)=U(|L_{x}-x|)$. In view
of this, it is easy to show that $F(k_{j_{1}},x_{2})$ is independent
of $x_{2}$: $F(k_{j_{1}},x_{2})=F(k_{j_{1}},0)$. As a result,
\begin{equation}
 \nu_{2}(k_{j_{1}},k_{j_{2}}) = L_{x}\nu^{c}(k_{j_{1}})\delta_{k_{j_{1}},-k_{j_{2}}},
    \label{11} \end{equation}
 \begin{equation}
\nu^{c}(k_{j_{1}})=F(k_{j_{1}},0),
    \label{11b} \end{equation}
\begin{equation}
 U(|x_{1}-x_{2}|) = \frac{1}{L_{x}}
 \sum\limits_{k_{j}}^{(2\pi)}\nu^{c}(k_{j})e^{ik_{j}(x_{1}-x_{2})},
     \label{12} \end{equation}
where $\delta_{k_{j_{1}},-k_{j_{2}}}$ is the Kronecker symbol.
We note that the particle on a ring undergoes the action of another particle from both sides. Therefore,
\begin{equation}
 U(|x_{1}-x_{2}|) = U_{1}(|x_{1}-x_{2}|)+U_{1}(L_{x}-|x_{1}-x_{2}|),
     \label{13} \end{equation}
where $U_{1}$ is the potential of such a force. If we unlock the
ring, then relation (\ref{13}) will contain only the first term. It
follows from (\ref{13}) that
$U(|x_{1}-x_{2}|)=U(L_{x}-|x_{1}-x_{2}|)$, which agrees with
(\ref{10}). For the potential $U_{1}(|x_{1}-x_{2}|)$ we have
\begin{equation}
F(k_{j},0) = \int\limits_{0}^{L_{x}}dx
U_{1}(|x|)e^{-ik_{j}x}=\nu(k_{j}).
    \label{14} \end{equation}
Let us consider $U_{1}(L_{x}-|x_{1}-x_{2}|).$ Using the change
$L_{x}-|x|=L_{x}-x = \tilde{x}$ with regard for $e^{ik_{j}L_{x}}=1,$
we obtain $F(k_{j},0)=\nu(-k_{j})$. Then
\begin{eqnarray}
&&\nu^{c}(k_{j}) = \nu(k_{j})+ \nu(-k_{j})=\label{15} \\
&=& \int\limits_{0}^{L_{x}}dx
U_{1}(|x|)(e^{-ik_{j}x}+e^{ik_{j}x})= \int\limits_{-L_{x}}^{L_{x}}dx
U_{1}(|x|)e^{-ik_{j}x}.
   \nonumber  \end{eqnarray}
Formulae (\ref{12}), (\ref{13}), and (\ref{15}) set the
Fourier-expansion of the potential for a cyclic 1D system. If we
turn $L_{x}$ to infinity  and neglect the image, then we obtain
T-expansion (\ref{1}), (\ref{2}) in 1D. This is the proof of the
T-expansion.

Note  that at the transition to the T-limit the potential
$U_{1}(L_{x}-|x_{1}-x_{2}|)$ (image) is usually neglected in
(\ref{13}). However, this is not entirely correct, because the
topologies of an infinite closed line and an infinite unclosed one
are different. The physical potentials $U_{1}(|x|)$ are usually
large at small $|x|$ and small at large $|x|$. If we turn
$L_{x}\rightarrow \infty$ in $U_{1}(L_{x}-|x_{1}-x_{2}|)$, we can
also turn $x_{1}$ to infinity so that the difference
$L_{x}-|x_{1}-x_{2}|$ be small. Then the potential
$U_{1}(L_{x}-|x_{1}-x_{2}|)$ is not small and, strictly speaking,
should not be neglected.

For a cyclic system, the potential $U(|x_{1}-x_{2}|)$ can be
expanded in a different way, by considering $|x_{1}-x_{2}|$ or
$x_{1}-x_{2}$ as an argument. In this case, it is necessary to take
the image into account (see (\ref{13})). At the end of this work, we
will show that the second way leads again to formulae (\ref{12}),
(\ref{13}), and (\ref{15}). Formulae (\ref{12}), (\ref{13}), and
(\ref{15}) can easily be generalized to three dimensions, in which
case the potential has $2^{3}-1=7$ images.

Let us return to the bounded system. Expanding in series, we
consider the arguments $x_{1}$ and $x_{2}$ to be independent. Then
the exact formulae (\ref{6}) and (\ref{7}) are valid. The system is
noncyclic, and, therefore, the potential has no property (\ref{10}).
Instead of (\ref{13}), we have
\begin{equation}
 U(|x_{1}-x_{2}|) = U_{1}(|x_{1}-x_{2}|),
     \label{21} \end{equation}
because no image is present. Consider the simple potential
(\ref{22}). Taking into account its image for a cyclic system, we
obtain
 \begin{equation}
\nu^{c}(k_{j_{1}}) = \frac{2U_{0}}{k_{j_{1}}}\sin{k_{j_{1}}a}.
    \label{23} \end{equation}
For a bounded system for $k_{j_{1}}\neq 0$ we get
\begin{equation}
 F_{b}(k_{j_{1}},x_{2}) =
\left [ \begin{array}{ccc}
    F_{1}(k_{j_{1}},x_{2}),  & \   0\leq x_{2}< a  & \\
    \nu^{c}(k_{j_{1}}),  & \   a< x_{2}< L_{x}-a  & \\
    F_{2}(k_{j_{1}},x_{2}),  & \ L_{x}-a <x\leq L_{x}, &
\label{24} \end{array} \right. \end{equation}
 \begin{eqnarray}
F_{1}(k_{j},x_{2}) &=& \frac{U_{0}}{k_{j}}(\sin{k_{j}a}+\sin{k_{j}x_{2}})+ \nonumber \\
&+&\frac{iU_{0}}{k_{j}}(\cos{k_{j}a}-\cos{k_{j}x_{2}}),
    \label{25} \end{eqnarray}
 \begin{eqnarray}
F_{2}(k_{j},x_{2})& =& \frac{U_{0}}{k_{j}}(\sin{k_{j}a}-\sin{k_{j}x_{2}})+ \nonumber \\
&+&\frac{iU_{0}}{k_{j}}(\cos{k_{j}x_{2}}-\cos{k_{j}a}),
    \label{26} \end{eqnarray}
\begin{equation}
 \nu_{2}(k_{j_{1}},k_{j_{2}}) = L_{x}\nu^{c}(k_{j_{1}})\delta_{k_{j_{1}},-k_{j_{2}}}+
 \tilde{\nu}_{2}(k_{j_{1}},k_{j_{2}}),
    \label{27} \end{equation}
\begin{equation}
 \tilde{\nu}_{2}(k_{j_{1}},k_{j_{2}}) = \int\limits_{0}^{a}
 \left [F_{1}(k_{j_{1}},x_{2})e^{-i(k_{j_{1}}+k_{j_{2}})x_{2}} + \mbox{c.c.}  \right ].
    \label{28} \end{equation}
Now the quantity $\nu_{2}(k_{j_{1}},k_{j_{2}})$ has a non-diagonal
part $\tilde{\nu}_{2}(k_{j_{1}},k_{j_{2}})\neq 0$, which is equal to
zero for a cyclic system. If we neglect it, we obtain
 T-expansion (\ref{1})-(\ref{3}). Thus, for an arbitrarily large but finite system with
boundaries, the T-expansion is not the exact Fourier series; it is
only an approximate expansion, which follows from the exact formulae
(\ref{6}), (\ref{27}) by neglecting $\tilde{\nu}_{2}$. To what
extent is this neglect proper? For the real-life systems we have $a
\ll L_{x},$ and the non-diagonal addition $\tilde{\nu}_{2}$ is
small. But sum (\ref{6}) contains much more non-diagonal terms, than
diagonal ones. The role of the non-diagonal contribution can be
easily estimated. Assume that it is insignificant and restore
potential (\ref{22}) with the help of the T-expansion. As a result,
we obtain (\ref{29}), which was mentioned above: the T-expansion
adds the image to the initial potential. If we do not reject
$\tilde{\nu}_{2}(k_{j_{1}},k_{j_{2}})$, then series (\ref{6}),
(\ref{27}) reproduces the initial potential exactly, without any
image.

With regard to the limit $L_{x}\rightarrow \infty$ one can say the
following. If we turn continuously $L_{x}\rightarrow \infty$ in the
formulae with a finite $L_{x}$, we always have a bounded system. In
other words, the limiting values for $L_{x}\rightarrow \infty$ are
the values for the arbitrarily large, but finite system. In this
case, it is necessary to consider
$\tilde{\nu}_{2}(k_{j_{1}},k_{j_{2}})$; otherwise, the image
appears. Thus, such procedure does not allow us to exactly obtain
the value for an infinite system without boundaries.

The following mathematical property is curious. It allows us to
better understand the role of images. Let us expand the potential
$U(L_{x}-|x_{1}-x_{2}|)$ of the image in the exact zero-series
(\ref{16}), (\ref{17}). For any $U(L_{x}-|x_{1}-x_{2}|),$ we obtain
the Fourier component
\begin{equation}
\nu_{im}(k_{j})=(-1)^{j}\nu_{z}(k_{j}),
\label{30}  \end{equation}
where $\nu_{z}(k_{j})$ is the Fourier-component of the potential  $U(|x_{1}-x_{2}|)$, and $k_{j}=\pi j/L_{x}$.
Then the Fourier-component of the zero-expansion of the potential $U(|x_{1}-x_{2}|)+U(L_{x}-|x_{1}-x_{2}|)$ reads
\begin{equation}
\tilde{\nu}(k_{j}) =
\left [ \begin{array}{ccc}
           0,  & j=2l+1, & \\
    2\nu_{z}(k_{j}),  & \   j=2l.  &
\label{31} \end{array} \right. \end{equation} Substituting it in
(\ref{16}), we obtain the T-expansion: formula (\ref{1}) with
$\nu(k)$ (\ref{2}). That is, \textit{the T-expansion for a system
with boundaries follows from the exact zero-series (\ref{16}),
(\ref{17}) with regard for images that are really lacking in the
system}. In this case, the account for the image  strongly modifies
the Fourier components of the potential: half of them become zero
and the other half is doubled. Formulae (\ref{30}), (\ref{31}) prove
also the above statement that the expansion of a potential in a
Fourier series for a cyclic system gives formulae (\ref{12}),
(\ref{13}), and (\ref{15}), if $x_{1}-x_{2}$ is taken as the
argument of the function.

\section{Physical consequences}
The above analysis implies that the T-expansion is not exact for a
bounded system. But it is usually applied namely to bounded systems.
The inaccuracy consists in that the T-expansion of the potential
$U(|x_{1}-x_{2}|)$ yields the potential
$U(|x_{1}-x_{2}|)+U(L_{x}-|x_{1}-x_{2}|)$ (in 1D). For a finite
cyclic system, this feature will give no error in the solution, if
the potential in a model is described by the right-hand side of the
formula (\ref{1}) (that is usually the case), since the right-hand
side of (\ref{1}) gives the complete cyclic potential
$U(|x_{1}-x_{2}|)+U(L_{x}-|x_{1}-x_{2}|)$. But the situation is
different for a bounded system. The formula (\ref{1}) gives
$U(|x_{1}-x_{2}|)+U(L_{x}-|x_{1}-x_{2}|)$, but there is no potential
$U(L_{x}-|x_{1}-x_{2}|)$ in the system, and the transition from
$U(|x_{1}-x_{2}|)$ to $U(|x_{1}-x_{2}|)+U(L_{x}-|x_{1}-x_{2}|)$
means the transition from a noncyclic complete potential of the
system to a cyclic potential. This changes the topology of the
system. Could this lead to an incorrect description of the
many-particle system? The examples (A) and (B) above show that in
the case of a long-range potential the answer is yes, but for a
short-range potential rather no.

Let us consider the second case --- a many-particle system with a
short-range interatomic potential.  In this case, the indicated
inaccuracy is small in the meaning that it concerns mainly atoms
near the walls of a vessel (the image $U(L_{x}-|x_{1}-x_{2}|)$ is
significant only in the case where one atom is located near one
wall, and another atom is positioned near the opposite wall). But
this inaccuracy is large in the meaning that it changes the topology
of the whole interaction in the system, by making the system closed
by the interaction. If the physics of the system is defined by
separate atoms, this change should not manifest itself in the bulk
properties, since the atom after the collision with a wall quickly
``forgets'' this wall, by moving inward the vessel and colliding
with other atoms. Therefore, for classical systems, the use of the
T-limit should not lead to incorrect results.  But at very low
temperatures (quantum liquids, quantum crystals, and quantum gases)
the physics of the system is formed by collective oscillations.
These oscillations occur in bulk, but they are modulated by the
walls. In this way, the walls can, in principle, affect the bulk
properties. In other words, it cannot be excluded that the topology
of the system strongly influences the wave functions and frequencies
of the collective excitations and thus the bulk properties of the
system (such a possibility was not considered in \cite{peierls}).
However, calculations
\cite{mtgp2012,mtmethodbog,cazalilla2002,cazalilla2004} show that
the dispersion law of a 1D system of spinless bosons at
$T\rightarrow 0$ does not change when periodic boundary conditions
are replaced by zero ones, provided that the interatomic potential
is short-range. That is, changing the topology of such a quantum
system does not lead to a change in the bulk properties. This
statement also seems to be true for other many-particle quantum
systems. In this case, the transition to the T-limit is justified.

\section{Conclusion}
We have studied different Fourier expansions of an interatomic
potential $U(|\textbf{r}_{1}-\textbf{r}_{2}|)$ for a finite system.
Such an analysis is useful for the accurate description of real-life
many-particle systems, which are usually bounded.   The analysis has
shown that the zero-expansion, modulus-expansion, and a double
Fourier series reproduce the interatomic potential
$U(|\textbf{r}_{1}-\textbf{r}_{2}|)$ exactly. Of these three exact
expansions, the simplest and most suitable for modern methods is the
zero-series (\ref{16}), (\ref{17}). In this case, the usually used
T-expansion distorts the potential so that a cyclic potential is
obtained instead of the initial noncyclic one; this changes the
topology of a problem. If the many-particle system is periodic, the
T-expansion gives the correct description. If the many-particle
system is bounded, but the interatomic interaction is short-range,
the T-expansion also seems to provide an accurate description.
However, if the system is bounded and the interatomic potential is
long-range, then the use of the T-expansion must lead to an
incorrect description of the system.



\begin{thebibliography}{200}
\bibitem {bogol1946} N.N. Bogoliubov, in {\it Studies in Statistical
Mechanics, I}, ed. by G.E. Uhlenbeck and J. de Boer (North-Holland,
Amsterdam, 1962).
\bibitem {ruelle} D.~Ruelle, {\it Statistical Mechanics. Rigorous Results}
      (Benjamin, New York, 1969).
\bibitem {peierls} R.~Peierls, {\it Surprises in Theoretical Physics}
      (Princeton University Press, Princeton, New Jersey, 1979), ch 3.
\bibitem {mtgp2012} M. Tomchenko, arXiv:1211.1723v4 [cond-mat.quant-gas].
\bibitem {mtmethodbog}  M.D. Tomchenko, \textit{Ukr. J. Phys.} \textbf{64}, 250 (2019).  https://doi.org/10.15407/ujpe64.3.250
\bibitem {kurant1} R. Courant and D. Hilbert, {\it Methods of Mathematical Physics}, Vol. 1
       (Interscience, New York, 1949).
\bibitem {cazalilla2002}  M.A. Cazalilla, \textit{EPL} \textbf{59}, 793
(2002).  https://doi.org/10.1209/epl/i2002-00112-5
\bibitem {cazalilla2004}  M.A. Cazalilla,
 \textit{J.~Phys.~B: At. Mol. Opt. Phys.} \textbf{37}, S1 (2004).
https://doi.org/10.1088/0953-4075/37/7/051



\end{thebibliography}
 \end{document}